\documentclass[fleqn,usenatbib]{mnras}

\usepackage{newtxtext,newtxmath}

\usepackage[T1]{fontenc}

\DeclareRobustCommand{\VAN}[3]{#2}
\let\VANthebibliography\thebibliography
\def\thebibliography{\DeclareRobustCommand{\VAN}[3]{##3}\VANthebibliography}

\usepackage{graphicx}	
\usepackage{amsmath}	
\usepackage{wasysym}    
\usepackage{bm}         

\usepackage{longtable}
\usepackage{multirow}
\usepackage{threeparttable}
\usepackage{paralist}
\usepackage{booktabs}
\usepackage{xcolor}
\usepackage{ulem}

\newcommand\Msun{\text{M}_{\astrosun}} 

\makeatletter 
  \patchcmd{\NAT@citex}
    {\@citea\NAT@hyper@{%
      \NAT@nmfmt{\NAT@nm}%
      \hyper@natlinkbreak{\NAT@aysep\NAT@spacechar}{\@citeb\@extra@b@citeb}%
      \NAT@date}}
    {\@citea\NAT@nmfmt{\NAT@nm}%
    \NAT@aysep\NAT@spacechar\NAT@hyper@{\NAT@date}}{}{}

  \patchcmd{\NAT@citex}
    {\@citea\NAT@hyper@{%
      \NAT@nmfmt{\NAT@nm}%
      \hyper@natlinkbreak{\NAT@spacechar\NAT@@open\if*#1*\else#1\NAT@spacechar\fi}%
        {\@citeb\@extra@b@citeb}%
      \NAT@date}}
    {\@citea\NAT@nmfmt{\NAT@nm}%
    \NAT@spacechar\NAT@@open\if*#1*\else#1\NAT@spacechar\fi\NAT@hyper@{\NAT@date}}
    {}{}
\makeatother

\title[Diffuse Ly$\alpha$ emission]{Cosmic Ly$\boldsymbol{\alpha}$ emission from diffuse gas}

\author[Tsai et al.]{
Sung-Han Tsai,$^{1,2}$\thanks{E-mail: m1069003@gm.astro.ncu.edu.tw}
Ke-Jung Chen,$^{1,3}$
Aaron Smith$^{4}$
and
Yi-Kuan Chiang,$^{1}$
\\
$^{1}$Institute of Astronomy and Astrophysics, Academia Sinica, Taipei 106319, Taiwan\\
$^{2}$Department of Physics, National Taiwan University, Taipei 10617, Taiwan\\
$^{3}$Heidelberg Institute for Theoretical Studies, Schloss-Wolfsbrunnenweg 35, Heidelberg 69118,
Germany\\
$^{4}$Department of Physics, The University of Texas at Dallas, Richardson, TX 75080, USA\\
}

\date{Accepted XXX. Received YYY; in original form ZZZ}

\pubyear{2025}

\begin{document}
\label{firstpage}
\pagerange{\pageref{firstpage}--\pageref{lastpage}}
\maketitle

\begin{abstract}
The Ly$\alpha$ emission has emerged as a powerful tool for probing diffuse gas within the large-scale structure of the universe. In this paper, we investigate cosmic Ly$\alpha$ emission by post-processing cosmological simulations from \texttt{IllustrisTNG}  project. We first define the intergalactic medium (IGM) and the circumgalactic medium (CGM), which are collectively referred to as diffuse gas, and then examine their individual contributions to the cosmic Ly$\alpha$ signal. Our results show that IGM dominates Ly$\alpha$ emission in regions with $n_{\rm H} < 10^{-2}\ \rm cm^{3}$, while CGM becomes dominant at higher densities. By analyzing the redshift evolution of Ly$\alpha$ emission mechanisms, we find that collisional excitation consistently dominates throughout cosmic time. In contrast, recombination exhibits a sharp increase from $z=6$ to $z=5$ due to activation of the time-variable UV background in the simulation. 
Compared to the intensity mapping measurement, our predicted diffuse Ly$\alpha$ luminosity density is approximately 65 and 120 times lower than current estimates at $z=1$ and $z=0.3$, respectively. Although the diffuse Ly$\alpha$ signal remains below current detection thresholds, our findings suggest it becomes intrinsically stronger at higher redshifts due to enhanced gas density, elevated temperatures, and a stronger UV background. Upcoming 30-meter-class ground-based telescopes, such as the E-ELT, TMT, and GMT, along with the space-based telescope SPHEREx, will significantly enhance the detection of Ly$\alpha$ emission from cosmic diffuse gas at high-$z$ through stacking and cross-correlation techniques.

\end{abstract}




\section{Introduction}\label{sec:Introduction}

Modern cosmological simulations in the $\Lambda$CDM paradigm have successfully reproduced the formation of the large-scale structure (LSS) of the universe \citep{1980peebles, 1996bond, 2005springel, 2015wang, 2017mccarthy}, including complex processes that produce realistic galaxy populations and their environments \citep{Vogelsberger2020,Angulo2022}. A defining feature of this structure is the cosmic web, characterized by its sheets and filaments that act as fundamental building blocks, intricately linking clustered galaxies and extending across the vast cosmic expanse.
The hierarchy of filaments channels gas and dark matter to influence the formation and evolution of galaxies within clusters \citep{2014dubois, 2018kraljic, 2023hasan}. 

The gas bound within haloes is generally termed the circumgalactic medium (CGM) when surrounding galaxies, and the hot intracluster medium (ICM) in galaxy clusters, whereas gas outside haloes pervading inter-halo space is the intergalactic medium (IGM). Collectively these components hold vital information to understand the thermal and dynamical state of galaxy clusters and their environments \citep{Gunn1972,Bhringer2010, Mantz2017}. The ICM consists mainly of ionized hydrogen and helium and accounts for most of the baryonic material in galaxy clusters. This superheated plasma, with a temperature of $\rm 10^{7}$ to $\rm 10^{8}$\,K, emits strong X-ray radiation \citep{Sarazin1986, Tozzi2001}. Furthermore, the exchange of accretion from and winds to the CGM and IGM, provides insight into the regulatory interactions and feedback processes at play \citep{Furlanetto2005, McQuinn2016, Tumlinson2017, Peeples2019, 2024Tung}. Despite their crucial role, the extremely low density of the CGM and IGM (hereafter collectively referred to as diffuse gas) poses significant observational challenges \citep{Martin2014, Pakmor2020}. In particular, the existence of a significant neutral atomic and molecular CGM component implies that solving CGM questions requires synergistic optical/UV and sub-mm observations. Unfortunately, none of the current submillimeter telescopes achieve the necessary sensitivity for such observations \citep{Lee2024}. New strategies are therefore needed to unveil the properties of diffuse cosmic gas.

One promising approach to probe diffuse gas is via Lyman-alpha (Ly$\alpha$) emission. Specifically, even low-density hydrogen is capable of producing Ly$\alpha$ photons through recombination cascades following ionization and collisional excitation of neutral hydrogen by electron impacts \citep{1987hogan, 2005Cantalupo, 2012rosdahl}. However, if the only ionizing radiation source is the cosmic ultra-violet (UV) background, the resulting Ly$\alpha$ surface brightness is expected to be extremely faint preventing robust direct observation. The detectability of this faint emission significantly increases if the diffuse gas is situated near strong UV sources, such as quasars, star-forming galaxies, or supermassive black holes \citep{1993Charlot, 1996Steidel, 2005Cantalupo, 2019Umehata}. In these cases, the brightness of Ly$\alpha$ becomes more prominent due to additional ionization of dense substructures or resonant scattering of Ly$\alpha$ photons from these intense nearby sources, enhancing its observability, sometimes in the form of giant Ly$\alpha$ nebulae and fluorescent filaments around these sources. Various technologies have been employed to overcome these observational challenges even for truly diffuse and isolated regions. These include stacking subcubes of observational data to achieve higher cumulative levels of surface brightness sensitivity \citep{2018Gallego}, and using integral field unit (IFU) spectroscopic imaging to observe extended emission around high-redshift galaxies \citep{1991Hu, Hill2008, 2014Roche, Bacon2021}. These have revealed filamentary Ly$\alpha$ structures that likely trace gas accretion onto galaxies \citep{Martin2023}. Another approach is employing line intensity mapping to statistically measure the cumulative emission from all sources including faint unresolved ones by mapping the sky in a given line and analyzing the aggregate signal or its fluctuations \citep{2019Chiang, Renard2024}. In particular, intensity mapping in Ly$\alpha$ with ultraviolet or optical surveys via cross-correlation with known tracers offers a way to measure the cosmic Ly$\alpha$ background contributed by diffuse IGM/CGM gas that is otherwise too faint to detect directly.

Despite these advances, observational data on cosmic Ly$\alpha$ emission remain sparse, especially at high redshift. Several factors contribute to this scarcity, such as sky background noise, line-of-sight absorption by the IGM, and instrumental sensitivity limitations \citep{Tapken2007, Ouchi2019}. This poses challenges for robustly interpreting the evolution of the Ly$\alpha$ luminosity density across cosmic time. Current measurements are limited in number and carry large uncertainties, complicating attempts to distinguish between different evolutionary scenarios. Given these limitations, theoretical modeling provides a crucial complementary approach. By using cosmological hydrodynamic simulations, we can predict the Ly$\alpha$ emission from known physical processes and compare these to the existing measurements, thereby improving our understanding of the Ly$\alpha$ budget of the Universe.

In this paper, we investigate diffuse Ly$\alpha$ emission by analyzing the outputs of state-of-the-art cosmological simulations. In particular, we utilize the large-volume \texttt{IllustrisTNG} simulations to calculate Ly$\alpha$ emission from three distinct components: galaxies, the CGM, and the IGM. This allows us to quantify the contributions of diffuse gas (CGM+IGM) to the total Ly$\alpha$ emission as a function of redshift, and to directly compare the simulated Ly$\alpha$ luminosity density predictions to recent observational intensity mapping estimates, and explore the physical properties of the Ly$\alpha$ emission at high redshifts. Our work builds on previous theoretical studies that have employed these and similar simulations, for example focusing on Ly$\alpha$ emitters and haloes \citep{Gronke2017, Byrohl2021, Kimock2021}, large-scale clustering properties \citep{Behrens2018}, or prospects of detecting the cosmic web \citep{Furlanetto2003, Bertone2012, 2020Elias, 2021Witstok, Byrohl2023}. Here we specifically aim to assess the total Ly$\alpha$ output from diffuse gas in a cosmological volume and evaluate whether simulations can reconcile this with the observationally inferred Ly$\alpha$-visible budget, and by extension constraints on the local UV background that ionizes the diffuse gas.

This paper is organized as follows. In Section 2, we describe the simulations and our methodology for identifying diffuse gas and computing its Ly$\alpha$ emission. In Section 3, we present our results for the physical properties of diffuse gas and its Ly$\alpha$ luminosity density as a function of redshift and compare them with observational data. In Section 4, we discuss our findings in context, comparing to previous studies, examining the robustness of our results under different assumptions, and considering observational implications for current and future Ly$\alpha$ surveys. Finally, we summarize our conclusions in Section 5, highlighting while the simulated signal is faint, upcoming efforts promise to eventually observing this cosmic glow.

\section{Methods}\label{sec:Method}


\subsection{\texttt{IllustrisTNG} simulations}

Our analysis utilizes large-volume cosmological hydrodynamic simulations from the \texttt{IllustrisTNG} project, which consists of volumes from $(50\,\text{cMpc})^3$ up to $(300\,\text{cMpc})^3$ evolved from early times $z \sim 127$ to the present day ($z=0$) with a comprehensive model of galaxy formation physics \citep{2018Pillepich, 2018Nelson, 2018Naiman, 2018Springel, 2018Marinacci, 2019Nelson, 2019Pillepich}. \texttt{IllustrisTNG} and other similar high-fidelity simulation suites deepen our understanding of galaxy formation in the context of cosmic structure evolution, facilitating comparisons with galaxy observations. \texttt{IllustrisTNG} was run with the moving-mesh code \texttt{AREPO} \citep{2010Springel}, which uses a quasi-Lagrangian unstructured Voronoi mesh to solve the magneto-hydrodynamics \citep[MHD;][]{Pakmor2011,Pakmor2013} equations via a second-order finite-volume method. Gravitational forces are calculated with an adaptive tree-particle-mesh algorithm \citep{Xu1995, Bagla2002, Bode2003}, which combines the particle-mesh method on large scales with a tree code to handle particle-particle interactions at small separations.
The suite systematically includes different physical box sizes, mass resolutions, and subgrid physics.
The workhorse in the \texttt{IllustrisTNG} simulations is the galaxy formation module \citep{2017Weinberger, 2018Pillepich_gf} based on the \texttt{Illustris} model \citep{Vogelsberger2013}, and we summarize the most relevant ingredients in this section.

\textbf{Galaxy formation subgrid physics:} Because the resolution of cosmological volumes is limited (gas cell sizes are kiloparsec-scale in TNG100), \texttt{IllustrisTNG} employs sub-resolution models for processes occurring on smaller scales, such as the interstellar medium (ISM) structure and stellar feedback. Star formation is modeled with an effective equation of state approach \citep{Springel2003}, wherein gas above a threshold hydrogen number density ($n_\text{H} \approx 0.13\,\text{cm}^{-3}$) is converted to stars stochastically following a Kennicutt--Schmidt relation with a Chabrier initial mass function \citep{Chabrier2003}. Stellar evolution and feedback are included: newly formed stars inject energy, mass, and heavy elements into surrounding gas based on tabulated population synthesis yields for core-collapse and Type Ia supernovae and asymptotic giant branch (AGB) winds, tracking nine elements (H, He, C, N, O, Ne, Mg, Si, Fe). In addition to the equilibrium primordial thermochemistry, metal-dependent radiative cooling is implemented based on the density, temperature, metallicity, and redshift of the ISM gas \citep{Smith2008, Wiersma2009}, allowing gas to cool to $T\sim10^4$\,K. The primary radiation source for is the background UV radiation field for cosmic reionization and photoionization heating, which is activated at $z=6$ and treated as spatially uniform, with corrections applied for partial self-shielding in dense regions \citep{1992Katz, 2009Faucher, 2013Rahmati}. The stellar feedback associated with star formation drives galactic outflows via hydrodynamically-decoupled wind particles launched directly from star-forming gas, exhibiting an assigned wind velocity corresponding to the local dark-matter velocity dispersion \citep{Vogelsberger2013, Torrey2014}. Wind particles inherit the physical properties of the gas in which they are launched, including their thermal energy and metallicity.

Additionally, supermassive black hole (SMBH) formation and feedback are modeled as seed black holes accrete gas and release energy via two modes, a high-accretion quasar mode (thermal energy injection) and a low-accretion kinetic wind mode \citep{2017Weinberger}. This is inspired by theoretical arguments regarding the inflow/outflow solutions of advection-dominated accretion flows within this regime \citep[see][]{Yuan2014}. This AGN feedback can heat and expel gas from haloes, significantly influencing the state of the CGM. Furthermore, regions near active galactic nuclei (AGN) experience modified cooling properties due to the additional radiation emitted by AGNs \citep{Vogelsberger2013}. All these subgrid prescriptions have been calibrated against a range of observations (galaxy stellar masses, star formation rates, halo gas content, etc.) and produce a population of galaxies and gaseous environments broadly consistent with the real Universe \citep{2019Nelson_a, Vogelsberger2020}. However, uncertainties remain in the subgrid models, e.g. the strength and timing of feedback or the small-scale structure of the ISM and CGM, that impact the predicted Ly$\alpha$ emission. 

\textbf{Simulation selection:} For our study, we primarily use the TNG100-1 simulation run, the highest-resolution $(110.7\,\text{cMpc})^3$ box, as it provides a large enough volume to capture representative statistics. TNG100-1 has a gas cell mass resolution of $1.4 \times 10^6\,\Msun$ and dark matter mass resolution of $7.5 \times 10^6\,\Msun$, with gravitational softening lengths for dark matter and stars of $0.7\,\text{kpc}$ at $z=0$, and a minimum adaptive softening for gas of $185\,\text{cpc}$.
These datasets include 14 full snapshots, covering a redshift range of $0 < z < 6$, allowing us to examine the evolution of cosmic Ly$\alpha$ emission from diffuse gas. We also evaluate the impact of simulation resolution on Ly$\alpha$ by comparing the results from TNG100-1, TNG100-2, and TNG100-3.

The \texttt{IllustrisTNG} simulations do not explicitly model on-the-fly radiative transfer, so the assumption of photoionization equilibrium with an approximate self-shielding correction is an important caveat simplifying the attenuation of the UV background and local sources in high-density gas. The impact of this approximation on Ly$\alpha$ emissivity in dense CGM regions will be discussed later. To better model the high-redshift Universe we incorporate data from the \texttt{THESAN} project \citep{Kannan2022, Garaldi2022, Smith2022}, which is a suite of large-volume radiation-magneto-hydrodynamic simulations that couple the \texttt{IllustrisTNG} galaxy formation physics with an on-the-fly radiative transfer solver for ionizing radiation \citep{Kannan2019}. \texttt{THESAN} explicitly tracks the inhomogeneous reionization of the IGM and self-shielding within the CGM rather than assuming a uniform UV background. It also includes an empirical dust model and additional ionizing photon sources from binary stars, making it well-suited to study Ly$\alpha$ emission mechanisms. We use the \texttt{THESAN}-1 run, which has a box size of $(95.5\,\text{cMpc})^3$, comparable to TNG100, but with higher resolution with gas masses of $\approx 5.82 \times 10^5\,\Msun$. For this work, we use snapshots in the range $5.5 < z < 6$ in our high-redshift comparisons, as \texttt{THESAN} provides a more realistic ionization state of the CGM and IGM than \texttt{IllustrisTNG}. Our results serve as a consistency check at overlapping redshifts, to increase our confidence at lower redshifts. Still, the two simulations employ different physics which could lead to potential systematic uncertainties in our analysis.

\subsection{Identifying diffuse gas (IGM and CGM)}
Our goal is to isolate gas in the simulations that belongs to the diffuse cosmic component, i.e. the IGM and CGM, rather than dense gas in galaxies. To accomplish this, we utilize the halo finding algorithms applied to the simulation data \citep{2001Springel}. The Friends-of-Friends (FoF) algorithm is first used on the dark matter particle distribution to identify halos (bound structures) by linking particles within a certain distance. A linking length of 0.2 times the mean interparticle separation is used, producing a catalog of dark matter halos representing from massive clusters down to small isolated galaxies. Afterwards, the Subfind algorithm is run on each FoF group to identify gravitationally self-bound substructures, effectively locating individual galaxies (subhalos) within a larger halo (galaxy cluster). Gas and star particles are assigned to these subhalos or remain attached to the parent FoF halo if not bound to a subhalo.

Using these catalogs, we classify each gas cell into one of three categories:
\begin{enumerate}
    \item \textit{Galactic gas:} Gas that resides in the inner region of a halo, associated with a galaxy. Operationally, we define this as any gas that is either part of a Subfind-identified subhalo, within the central 10\% of the virial radius for the central galaxy, or in the effective equation of state identified as eligible for star formation anywhere in the halo, i.e. $\text{SFR} >0$. This includes dense ISM gas and immediate circumstellar regions that are not part of the diffuse medium.
    \item \textit{Circumgalactic medium (CGM):} Gas within a halo (FoF group) but not counted as galactic gas. In practice, this is gas that is not gravitationally bound to any subhalo (referred to as the ``inner fuzz'') in the outskirts of subhalos beyond $0.1\,R_\text{vir}$ of the halo. The CGM thus represents the halo gas envelope surrounding galaxies. We exclude any star-forming gas as these dense clumps are better thought of as part of satellite galaxies or local dense clouds.
    \item \textit{Intergalactic medium (IGM):} Gas that is not associated with any FoF halo, i.e. the low-density regions between halos (referred to as the ``outer fuzz''). The IGM fills the cosmic web filaments, sheets, and voids outside virialized structures.
\end{enumerate}
With these definitions, we consider ``diffuse gas'' as CGM + IGM in our analysis of cosmic Ly$\alpha$ emission, and compare these to the contribution from galaxies. We note that the exact boundary between ``galaxy'' and ``CGM'' is somewhat arbitrary and mostly meant to exclude the brightest central region. Other studies sometimes use different criteria or define the IGM by a fixed overdensity threshold. We will later discuss how such choices can affect the inferred Ly$\alpha$ emission.

\begin{figure*}
\centering
\includegraphics[width=.8\textwidth]{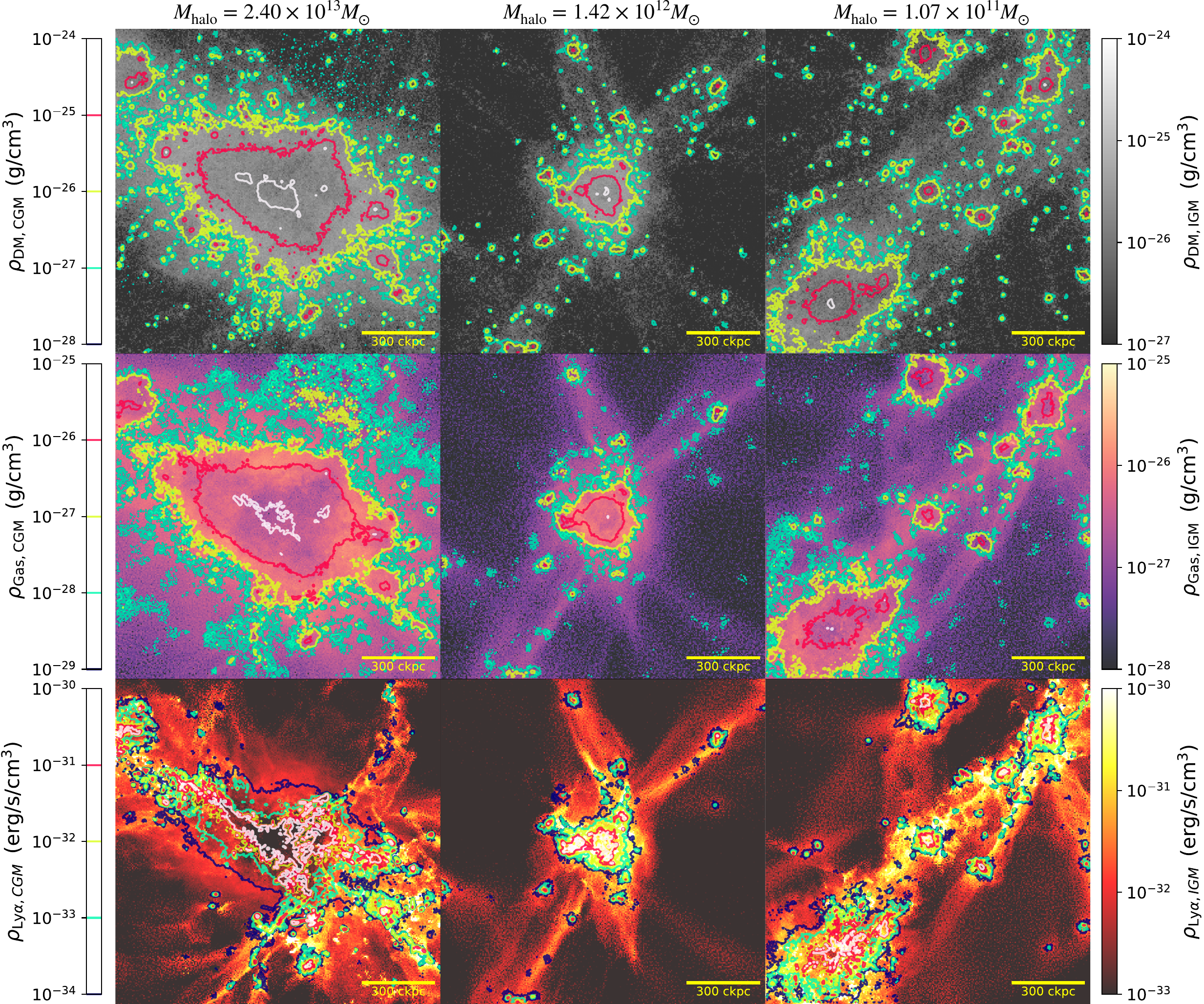}
\caption{Spatial distribution of dark matter, gas, and Ly$\alpha$ emission (top to bottom) within $2\,\text{cMpc}/h$ cubes centered on a halo at $z = 1$. From left to right, the central halo masses are $2.40 \times 10^{13}\,\Msun$, $1.42 \times 10^{12}\,\Msun$, and $1.07 \times 10^{11}\,\Msun$, respectively. The IGM (colormap) connects halos across cosmic structures, while the CGM (contours) closely surrounds galaxies. We define diffuse gas as the sum of the IGM and CGM.}
\label{fig1}
\end{figure*}

To visualize these components, in Figure~\ref{fig1} we show representative slices within a cubic region of $2\,\text{cMpc}/h$ per side at $z = 1$ in the simulated volume, including spatial distributions of dark matter density, gas density, and Ly$\alpha$ emissivity (from top to bottom). Each panel shows the IGM by a colour map and the CGM regions by contour levels. From left to right, the panels are centered on a halo with a dark matter mass of approximately $10^{13}$, $10^{12}$, and $10^{11}\,\Msun$, respectively. These selected regions include contributions from the central host halo and other surrounding halos. These illustrate that the IGM forms a filamentary network connecting halos, with relatively low densities and faint emission, while the CGM concentrates around halo centers with higher gas density. The morphology of strong Ly$\alpha$ surface brightness regions are more pronounced than the dark matter and gas distributions. Ly$\alpha$ emission is enhanced in denser regions by two-body $\rho^2$ biasing, and because the CGM is more shock-heated (leading to more collisional excitations) compared to IGM voids.

The qualitative picture reinforces that both components (CGM and IGM) are needed to describe the full diffuse Ly$\alpha$ background, and their contributions may vary with environment. For example, there is brighter and more extended CGM emission around a massive cluster than isolated galaxies. Intermediate environments contain several smaller halos within the cosmic web, revealing how diffuse gas bridges structures on these scales. These figures collectively demonstrate that the role of diffuse gas varies significantly depending on the mass of the central halo, influencing the structure of the surrounding IGM and CGM.

\subsection{Calculating Ly\texorpdfstring{$\boldsymbol{\alpha}$}{α} emission}

We compute the Ly$\alpha$ luminosity produced by each gas cell from the simulation outputs in post-processing assuming the production is through collisional excitation of neutral hydrogen and recombination of ionized hydrogen \citep{2019Aaron}. For simplicity, we ignore scattering and focus on the global budget, including an empirical correction for dust absorption within dense regions.

The collisional excitation emissivity is computed as follows. When a free electron with number density $n_e$ collides with a neutral hydrogen atom (number density $n_\text{HI}$), it can excite the atom from the ground state to the $2p$ state. The subsequent decay from $2p$ back to $1s$ emits a Ly$\alpha$ photon (wavelength 1216\,\AA). The collisional Ly$\alpha$ luminosity can be written as:
\begin{equation} \label{eq1}
  L_{\alpha}^\text{col} = h\nu_{\alpha} \int q_\text{1s2p}(T) n_e n_\text{HI}\,\text{d}V \, ,
\end{equation}
where the energy of a Ly$\alpha$ photon is $h\nu_\alpha = 10.2\,\text{eV}$ and the rate coefficient $q_\text{1s2p}(T)$ is taken from \citet{1991Scholz}. The collisional emissivity is highly sensitive to temperature, peaking around $T \sim 10^5$--$10^6$~K when a significant fraction of hydrogen is still neutral but electrons have enough kinetic energy to excite the Ly$\alpha$ transition efficiently.

The recombination emissivity accounts for free protons capturing an electron that cascades down to the ground state. The Ly$\alpha$ luminosity from recombinations can be expressed as:
\begin{equation}\label{eq2}
L_{\alpha}^\text{rec} = h\nu_{\alpha} \int P_\text{B}(T) \alpha_\text{B}(T) n_e n_\text{HII}\,\text{d}V \, ,
\end{equation}
where $n_\text{HII}$ is the proton number density, and $P_{\rm B}(T)$ represents the Ly$\alpha$ conversion probability per recombination event. For Case B, where the gas temperature is approximately $10^4$\,K, the $P_{\rm B}(T)$ is around $0.68$ although we adopt the fit from \citet{2008Cantalupo}. $\alpha_{\rm B}$ denotes the Case B recombination coefficient taken from \citet{Hui1997}. Case B assumes that Lyman continuum photons emitted by direct recombination to the ground state are locally absorbed by ionizing another nearby atom. In extremely low-density diffuse gas Case A (no trapping of Lyman continuum) might be more accurate, but this difference is a relatively minor caveat in the temperature range of interest.

In order to compare the contribution between the diffuse gas and galaxy, we also calculate the Ly$\alpha$ emission from stars as part of the galaxy contribution. This nebular component from simulations like \texttt{IllustrisTNG} without on-the-fly ionizing radiative transfer can be accounted for in one of two ways, either by conversion from the star-formation rate (SFR) or directly from the star particles. We choose the latter but first discuss the former as it is employed by other works \citep[e.g.][]{Byrohl2021}. Gas cells with non-zero SFR are part of the ISM of galaxies. The simulation treats such gas as a multi-phase medium where cold clouds and hot ionized phases coexist \citep{Springel2003}. In reality, this gas is comprised of a complex distribution of molecular star-forming clouds with young compact HII regions and older disrupted ones. However, the exact ionization states and unresolved collisional excitation and recombination emission cannot be accurately recovered, therefore we do not include any $\text{SFR} > 0$ gas cells in the CGM/IGM or in our reported resolved gas emission components. Finally, it is possible to use the smooth distribution of $\text{SFR} > 0$ cells as an emission source adopting standard conversions to hydrogen-ionizing (Lyman continuum; LyC) photons that are assumed to be locally reprocessed into Ly$\alpha$ photons, but we do not adopt this approach.

Instead, we directly convert the LyC output from star particles in the simulation to Ly$\alpha$ photons. These unresolved HII regions are part of the galaxy contribution to the intrinsic Ly$\alpha$ budget. Specifically, the Ly$\alpha$ luminosity from each star is
\begin{equation}\label{eq3}
L^\text{stars}_{\alpha} = 0.68 h\nu_{\alpha} (1 - f^\text{ion}_\text{esc}) \dot{N}_\text{ion} \, .
\end{equation}
Here, 0.68 represents the fraction of recombinations that yield Ly$\alpha$ for gas at $10^4$\,K. $f^\text{ion}_\text{esc}$ denotes the escape fraction of ionizing photons that escape into the IGM without inducing local recombinations, which is highly uncertain but low enough that we take this to be zero \cite{Smith2022}. We obtain $\dot{N}_\text{ion}$ for each star particle using stellar population synthesis tables \citep[calculated using the BPASS v2.2.1 models;][]{Eldridge2017}, based on the age and metallicity of each star particle. These models provide a framework to interpret the integrated light from galaxies and stellar populations, including the effects of binary star systems.

In this work, to obtain the observed Ly$\alpha$ emission from the galaxy component (including the inner-CGM of $<0.1\,R_\text{vir}$), we further multiply the intrinsic luminosities by the redshift-dependent galaxy Ly$\alpha$ escape fraction based on the empirical fit by \citet{Hayes2011}, described by the following relation:
\begin{equation}\label{eq4}
f^{\text{Ly}\alpha}_\text{esc} = 2.2 \times 10^{-3} \times (1+z)^{2.45} \, .
\end{equation}
This fraction monotonically decreases as a result of the increasing dust content in galaxies from $z = 6$ to $z = 0$, accounting for absorption effects in dense regions of the universe. This $f^{\text{Ly}\alpha}_\text{esc}$ is applied uniformly to all galaxies at a given redshift to all non-diffuse components. While this is overly simplistic, it should be sufficient for global quantities as it was effectively calibrated to match observed Ly$\alpha$ luminosity densities of galaxies as a function of redshift.

We note that our results and component classifications might be affected by Ly$\alpha$ resonant scattering. However, in diffuse gas of low-to-moderate density (IGM and outer CGM) the precise position of last scattering should not impact our main conclusions, especially given the relatively high ionization fraction in the post-reionization era. In the dense ISM and inner CGM, resonant trapping is important but approximately captured through the empirical escape fraction treatment. A more sophisticated approach would be to perform full Ly$\alpha$ radiative transfer calculations on the gas distribution \citep[e.g. as done by][]{Gronke2017, Byrohl2021, Byrohl2023} but that is beyond our current scope. We instead include the caveat that our predicted Ly$\alpha$ luminosities from the galaxy and CGM components carry significant uncertainties. We will highlight where such uncertainties could explain differences between our results and observations.


\section{Results}

\subsection{Diffuse Ly\texorpdfstring{$\boldsymbol{\alpha}$}{α} properties}

\begin{figure*}
\centering
\includegraphics[width=1.\textwidth]{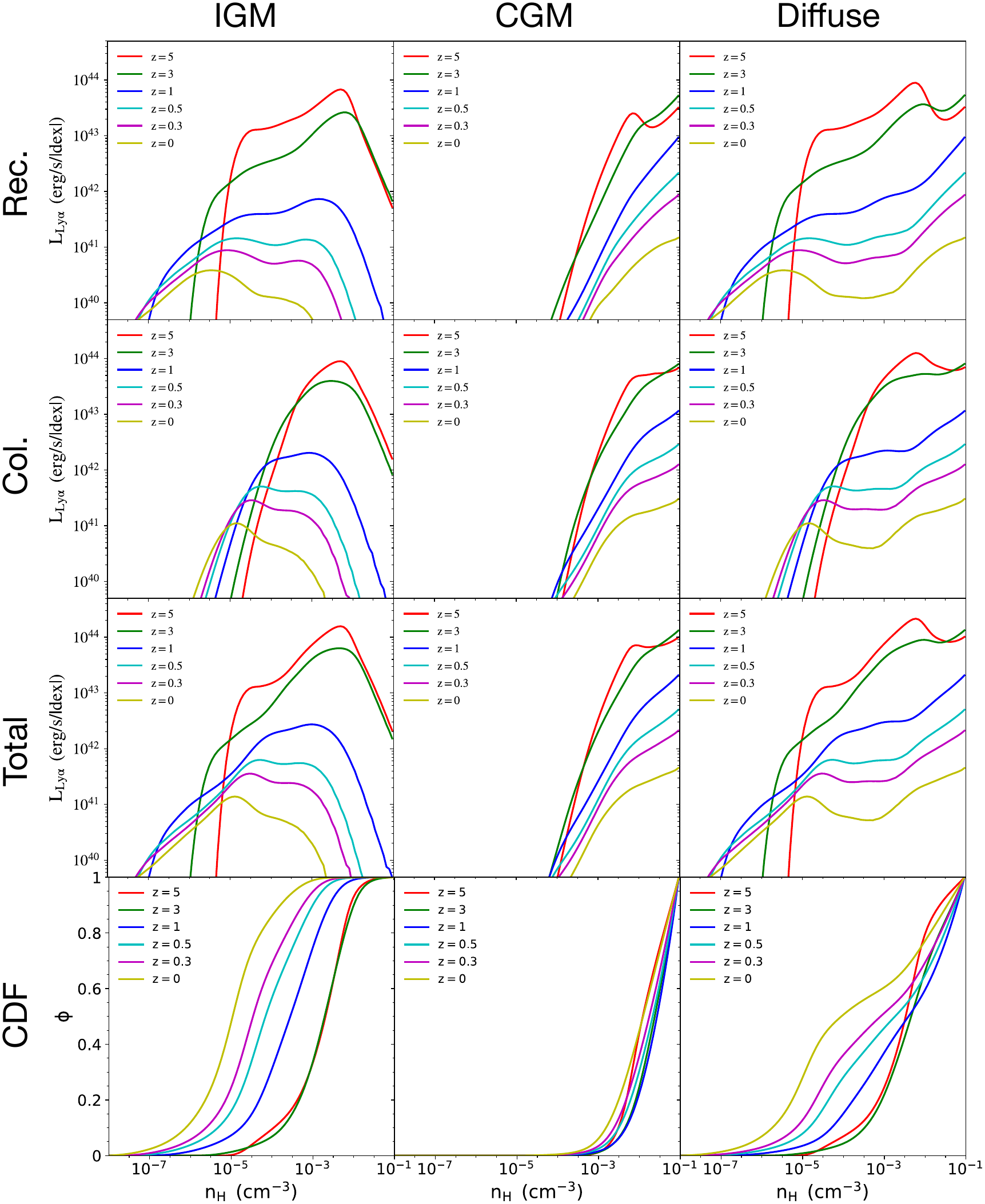}
\caption{Relationship between $n_{\rm H}$ and diffuse $\rho_{\rm Ly\alpha}$ at different redshifts. We separate the diffuse $\rho_{\rm Ly\alpha}$ into contributions from recombination ("Rec.") and collisional excitation ("Col.") processes within the IGM and CGM regions in top two row panels. The panels in the third row labeled "Total" show the combined luminosity density from both mechanism, where Total = Rec. + Col. The panels in the bottom row show their  cumulative distribution function (CDF), the $\rm \phi $ in the y-axis shows the cumulative fraction of the total Ly$\alpha$ luminosity contributed by gas with hydrogen number density below a given $n_{\rm H}$.}
\label{fig2}
\end{figure*}

We begin by examining the physical conditions of the diffuse gas and how they relate to Ly$\alpha$ emission. Figure~\ref{fig2} shows the Ly$\alpha$ luminosity ($\rm L_{\rm Ly_{\alpha}}$) as a function of gas density at representative redshifts for diffuse gas in the TNG100-1 simulation. We separate diffuse $\rm L_{\rm Ly\alpha}$ into contributions from recombination and collisional excitation processes within IGM and CGM to better understand their respective properties and evolution. The panels show that, for the IGM, $\rm L_{\rm Ly\alpha}$ increases with hydrogen number density to $n_{\rm H} \sim 10^{-2}\ \rm cm^{-3}$ and then decreases beyond this point. In contrast, the CGM component continues to increase steadily up to our maximum density bin at $n_{\rm H}=10^{-1}\ \rm cm^{-3}$. As a result, the diffuse $\rm L_{\rm Ly\alpha}$ exhibits a dominance transition: the regime with $n_{\rm H}<10^{-2}\ \rm cm^{-3}$ is dominated by IGM emission, while the highger density regime is increasingly governed by CGM emission. This density-dependent behavior reflects the different physical conditions and environments between the IGM and CGM. 

We further find a clear bifurcation in the dominant Ly$\alpha$ emission mechanism as a function of density. In regions with hydrogen number densities less than $10^{-5}\ \rm cm^{-3}$ ($n_{\rm H} \lesssim 10^{-5}\ \rm cm^{-3}$), recombination emission dominates over collisional emission. In these tenuous regions, the UV background can easily ionize the gas and produce more electrons and protons for recombination. As a result, recombination between protons and electrons is efficient, while the low-density gas suppresses collisional excitation. At higher densities ($n_{\rm H} \gtrsim 10^{-4}\ \rm cm^{-3}$), collisional excitation becomes the primary mechanism. Gas in these regions, typically dense filaments or the CGM, has been shock heated or photo heated to $T \sim 10^4$–$10^5$~K, producing a substantial population of thermal electrons that excite the remaining neutral hydrogen atoms. Although the neutral fraction is low, combining high electron density and sufficient residual neutrals leads to strong collisional Ly$\alpha$ emission. The CGM near halo centers generally lies in this intermediate-to-high density and temperature regime, thus contributing most of the diffuse Ly$\alpha$ luminosity density. For example, as shown in Figure~\ref{fig2}, CGM gas at $n_{\rm H} \sim 10^{-3}$–$10^{-1}\ \rm cm^{-3}$ dominates the total Ly$\alpha$ output, whereas IGM gas dominates the output with $n_{\rm H} < 10^{-3}\ \rm cm^{-3}$.

A particularly notable feature is the sharp drop in the emission driven by recombination Ly$\alpha$ between $z=3$ and $z=1$, observed in both the IGM and CGM regimes. This decline can be attributed not only to the decrease in gas density due to cosmic expansion but also to the weakening of the UV background. In the \texttt{TNG} simulations, the UV background peaks around $z\sim2-3$ and declines rapidly at lower redshifts. As the ionizing background weakens, the ionization fraction decreases, reducing the abundance of H II and free electrons required for recombination, and diminishing the collisional excitation efficiency as the supply of free electrons drops. Consequently, Ly$\alpha$ emission becomes less efficient, especially in the IGM, where gas densities are lower, and self-shielding is less effective. In contrast, the CGM maintains higher gas densities and stronger self-shielding, which help support a partially ionized state and preserve the conditions necessary for both recombination and collisional excitation, leading to a slower decrease in Ly$\alpha$ emission.

The cumulative distribution function (CDF) panels show the cumulative Ly$\alpha$ luminosity as a function of gas density threshold at several redshifts. At high redshift (e.g., $z = 6$), the Ly$\alpha$ emission is concentrated in relatively high-density gas, primarily originating from collisional excitation in shock-heated regions. As cosmic time progresses, the distribution of Ly$\alpha$ emitting gas broadens toward lower densities: by $z = 0$, a non-negligible fraction of the Ly$\alpha$ luminosity originates from gas as diffuse as $n_{\rm H} \sim 10^{-6}$--$10^{-5}\ \rm cm^{-3}$. This trend reflects the combined effects of cosmic expansion, which lowers the mean gas density, and a decaying clustering effect of cosmic structure. At $z \sim 1$, the evolution of large-scale become more steady and rates of halo assembly declines, reducing the formation of new dense regions. As a result, the IGM becomes more diffuse and less concentrated in filamentary structures, shifting the Ly$\alpha$ emitting regions toward lower density gas.

However, at all epochs, most of the Ly$\alpha$ luminosity still originates from overdense gas relative to the cosmic mean. Although the contribution from lower density regions increases at lower redshift, the higher density gas associated with the CGM continues to dominate. This underscores that the CGM, particularly the densest diffuse gas residing just outside galaxies, remains a crucial source of Ly$\alpha$ emission, even when focusing exclusively on the diffuse components.


\subsection{Gas phase diagrams}

\begin{figure*}
\centering
\includegraphics[width=1.\textwidth]{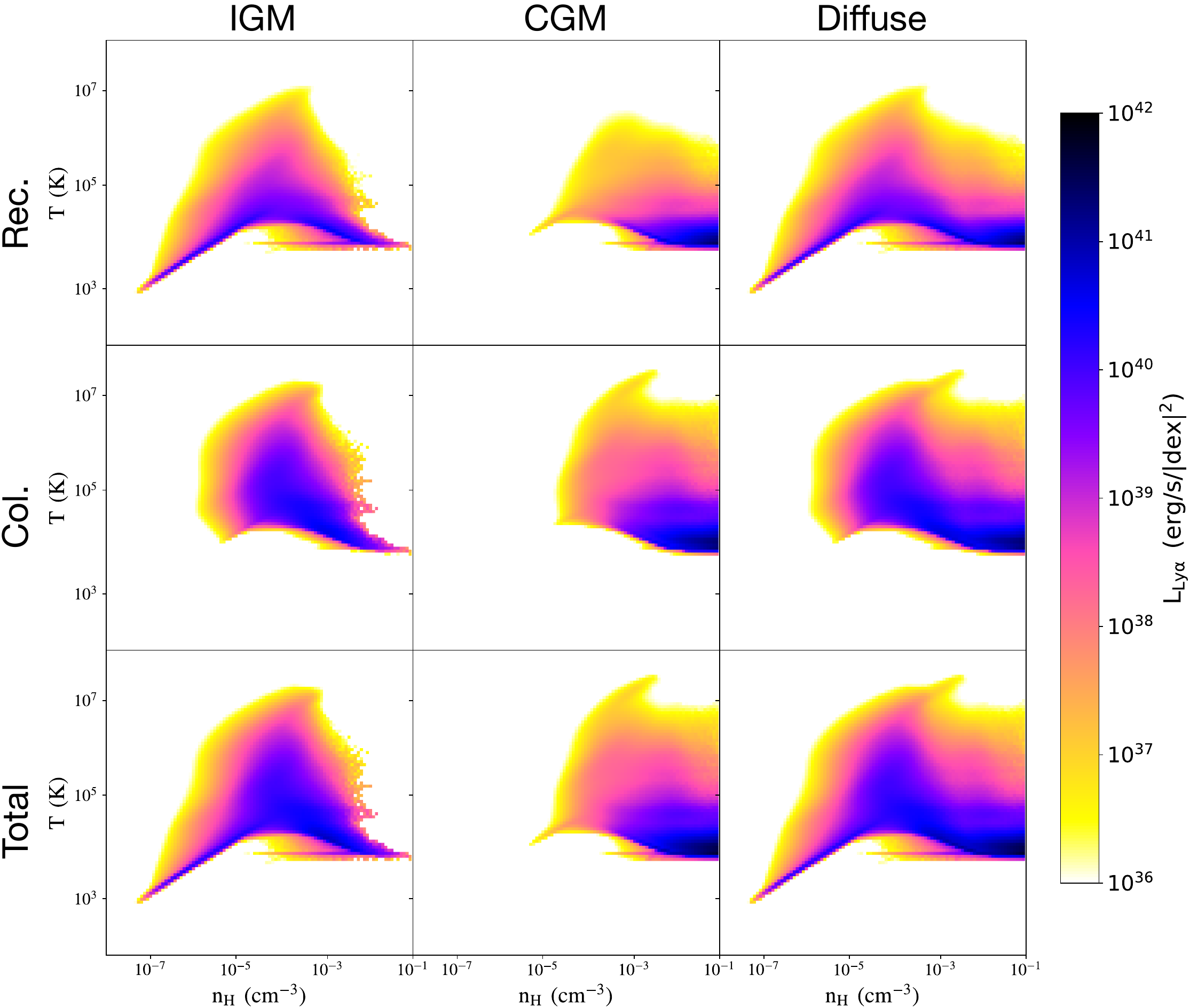}
\caption{Density-temperature-emission diagrams for diffuse gas at $z = 1$. From top to bottom, the panels show Ly$\alpha$ luminosity from recombination (Rec.), collisional excitation (Col.), and their total contribution (Total = Rec. + Col.). The panels represent the IGM, CGM, and diffuse gas components from left to right. Each panel consists of a $100 \times 100$ pixel grid}
\label{fig3}
\end{figure*}

\begin{figure*}
\centering
\includegraphics[width=1.\textwidth]{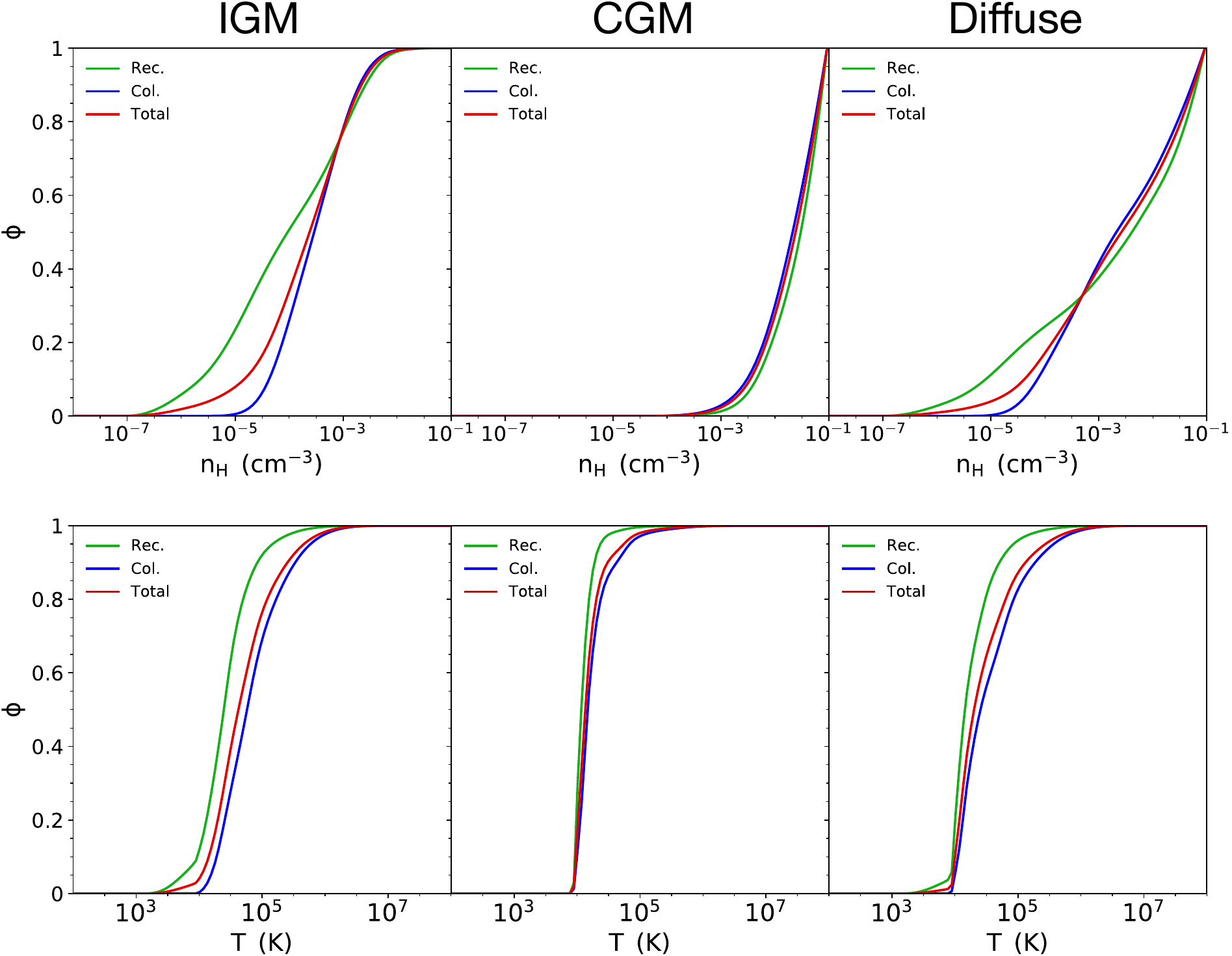}
\caption{CDFs of Ly$\alpha$ luminosity at $z=1$ as functions of $n_{\rm H}$ (top) and $T$ (bottom). From left to right, the panels correspond to IGM, CGM, or diffuse gas, with curves showing contributions from recombination, collisional excitation, and their sum.}
\label{fig4}
\end{figure*}

To better understand the Ly$\alpha$ emission from diffuse gas, we present the gas density-temperature-emission phase diagrams in Figure~\ref{fig3} and their corresponding CDF plots in Figure~\ref{fig4}. In IGM, Ly$\alpha$ emission arises primarily from gas at $T \sim 10^{4}-10^{5}\,\rm K$, where collisional excitation is the dominant emission mechanism. In this temperature range, free electrons possess sufficient kinetic energy to excite neutral hydrogen (H I) through collisions, resulting in efficient Ly$\alpha$ photon production. As the temperature increases beyond $10^{5}$ K, most of the neutral hydrogen becomes ionized, reducing the availability of H I for excitation and thus decreasing the collisional Ly$\alpha$ emission.
In addition to this main emission regime, the phase diagram reveals a tail in the recombination component that extends into low-density ($\rm n_{\rm H} < 10^{-5}$) and low-temperature ($T < 10^{4}\,\rm K$) regions. This tail probably originates from an ionized gas (H~II) that has cooled and begun to recombine in the underdense IGM, producing Ly$\alpha$ photons as it approaches thermal and ionization equilibrium. However, gas cooler than $10^4$ K generally emits much less Ly$\alpha$, since any neutral hydrogen in this regime produces 21-cm or fluorescent emission only in the presence of ionizing radiation. Moreover, even UV background ionized gas at $T \sim 10^4$ K has relatively low collisional excitation rates, limiting its Ly$\alpha$ output.
A noticeable horizontal feature near $T \sim 10^{4}\, \rm K$ appears in the phase diagrams. This corresponds to the transition in the gas cooling function: above $10^{4}\, \rm K$, line cooling processes are highly efficient, but below this temperature, the cooling rate drops sharply, causing the gas to linger at $10^{4}\, \rm K$. Additionally, this temperature marks the characteristic balance point where UV background photoheating cancels radiative cooling in low-density gas. As a result, the gas becomes thermally trapped around $10^{4}\, \rm K$, creating a plateau-like structure in the emission phase diagrams.

The CGM regions exhibit trends similar to those in the IGM, with Ly$\alpha$ emission arising mainly from gas at $T \sim 10^{4}-10^{5}\, \rm{K}$, where collisional excitation is the primary process. Compared to the IGM, the recombination Ly$\alpha$ emission is notably stronger in the CGM, due to its higher gas density, which promotes more frequent electron-H~I interactions. Although a horizontal structure around $T \sim 10^{4}\, \rm K$ is also present in the CGM phase space, it follows the same physical origin as in IGM, resulting from a combination of cooling inefficiency and thermal adjustment by the UV background.

Figure~\ref{fig3} shows that the emission Ly$\alpha$ originates mainly from the gas at $T \sim 10^{4}-10^{5} \rm{K}$, with collisional excitation dominating this regime. In contrast, recombination in the IGM displays a tail extending into the low-density, low-temperature region as ionized gas cools and begins to recombine. The CGM, because of its higher density, maintains a stronger recombination emission overall. The horizontal feature at $T \sim 10^{4}\, \rm K$ across both IGM and CGM reflects a common thermal stabilization process shaped by cooling properties and UV background photoheating.

Figure~\ref{fig4} presents the CDFs of Ly$\alpha$ luminosity with respect to hydrogen number density (top row) and temperature (bottom row) at $z=1$, for the IGM, CGM, and diffuse components. In the density panels, the CDFs show that for IGM, most Ly$\alpha$ photons are emitted by gas with $n_{\rm H} \lesssim 10^{-2}\ \rm cm^{-3}$. The CDF of the recombination in the IGM extends more smoothly into lower-density regions than the collisional, indicating that the recombination's emission is more evenly distributed in regions of different hydrogen number densities. In contrast, the CGM exhibits a steeper CDF, with Ly$\alpha$ emission concentrated at higher densities ($n_{\rm H} \gtrsim 10^{-3}\ \rm cm^{-3}$). The steeper rise in CGM CDFs reflects a more concentrated distribution of Ly$\alpha$ emission in higher-density regions, characteristic of gas residing near halos. These trends highlight that CGM emits more efficiently in dense environments, whereas the IGM emission is spread over a broader range of lower densities. In the temperature panels, most Ly$\alpha$ emission accumulates within the range $T \sim 10^{4}$--$10^{5.5}\ \rm K$, consistent with the regime where collisional excitation is most effective. The recombination CDFs in both the IGM and CGM show a broader spread toward lower temperatures, tracing recombining H~II in cooler, possibly recently cooled regions. The sharp increase in all curves near $T \sim 10^4\ \rm K$ further emphasize the importance of this characteristic temperature, where cooling becomes less efficient and UV background heating balances radiative losses, producing the plateau seen in the emission phase diagrams.

\subsection{Evolution of Ly\texorpdfstring{$\boldsymbol{\alpha}$}{α} emissions across cosmic time}
\begin{figure*}
\centering
\includegraphics[width=1.\textwidth]{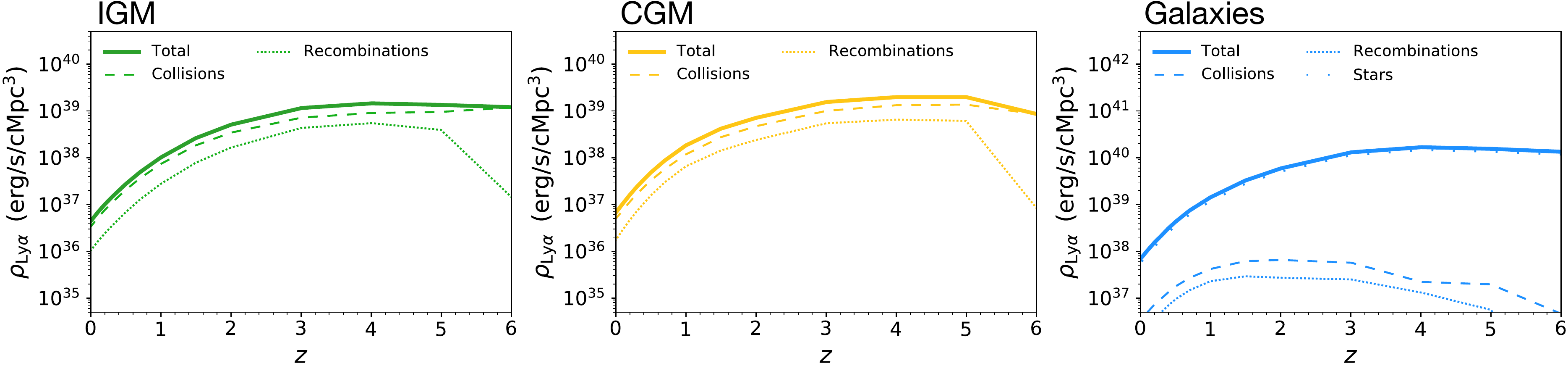}
\caption{The left and middle panels display Ly$\alpha$ emission from recombination and collisional excitation in the IGM and CGM, respectively. The right panel shows these contributions within galaxies, including the emission from stars.}
\label{fig5}
\end{figure*}

Figure~\ref{fig5} presents the redshift evolution of the Ly$\alpha$ luminosity density $\rho_{\rm Ly\alpha}$ from $z=6$ to $z=0$, separated into contributions from the IGM, CGM, and Galaxies. In general, collision excitation dominates over recombination in both IGM and CGM, typically contributing 2 to 3 times more Ly$\alpha$ luminosity. This implies that a non-negligible amount of H~I exists in these regions, allowing for efficient excitation by free electrons. The enhanced contribution from the CGM is expected, as gas in halos tends to reach higher densities and temperatures, increasing the local emissivity and making the CGM a more efficient Ly$\alpha$ emitter per unit volume. In the Galaxies panel, Ly$\alpha$ emission is overwhelmingly dominated by the stellar component at all redshifts. Stellar Ly$\alpha$ consistently contributes more than $90\%$ of the total luminosity in the central regions of halos, and the contributions from recombination and collisional excitation within the $SFR = 0$ gas are minimal in comparison.

A particularly notable feature in the IGM and CGM panels is the sharp increase in recombination Ly$\alpha$ luminosity density between $z=6$ and $z=5$. This is a direct consequence of the \texttt{TNG} simulation switching on the UV background at $z=6$, which starts the cosmic reionization and increases free electron density rapidly. Increasing free electrons can increase the rate of hydrogen recombination to produce more Ly$\alpha$ during $z=6$ to $z=5$. At $z \gtrsim 6$, collisional excitation overwhelmingly dominates the diffuse Ly$\alpha$ production: collisional processes generate over two orders of magnitude more Ly$\alpha$ luminosity than recombinations in diffuse gas. This occurs because during and just after reionization, diffuse gas (even in the IGM) is still being shock-heated and collisionally ionized, while the UV background remains weak, making photoionization and recombination relatively inefficient. At lower redshifts ($z < 5$), collisional excitation remains the dominant mechanism, but only by a factor of a few. Typically, collisional processes account for 60-70$\%$ of diffuse Ly$\alpha$ emission, while recombination contributes 30–40$\%$. The narrowing of this gap is driven by the growth of the UV background, which maintains ionization in the diffuse gas and increases recombination emission, and by the gradual decline of large-scale clustering at low redshift, which reduces shock heating and collisional excitation.

The diffuse $\rho_{\rm Ly\alpha}$ increases significantly at higher redshifts, peaking around $z=4$. This trend can be attributed to several physical effects: First, the gas density increases with redshift due to cosmological expansion, which raises the probability of collisional excitation and recombination events. Second, the gas temperatures are generally higher in the early universe, allowing more gas to reach or exceed the $T \sim 10^{4}\, \rm K$ threshold, where both collisional excitation and recombination become efficient. Third, the UV background is stronger at $z>2$, contributing to higher ionization fractions and, in turn, enhancing recombination Ly$\alpha$ emission in diffuse gas. Compared to the peak, the total Ly$\alpha$ luminosity density has dropped by about 2–3 orders of magnitude at $z=0$. This dramatic evolution is driven largely by the decline in cosmic star formation rate at late times (hence fewer ionizing photons and Ly$\alpha$ from galaxies) as well as reduction of diffuse gas.

Figure~\ref{fig5} also shows that $\rho_{\rm Ly\alpha}$ from collisional excitation and recombination in the Galaxies region is comparable to that of diffuse gas when $z<1$. However, at higher redshifts, the Ly$\alpha$ luminosity from collisional and recombination processes in the IGM and CGM can exceed that of the Galaxies by up to two orders of magnitude. This emphasizes the increasing role of diffuse baryons in producing observable Ly$\alpha$ emission during the earlier stages of cosmic evolution.

\begin{figure*}
\centering
\includegraphics[width=0.8\textwidth]{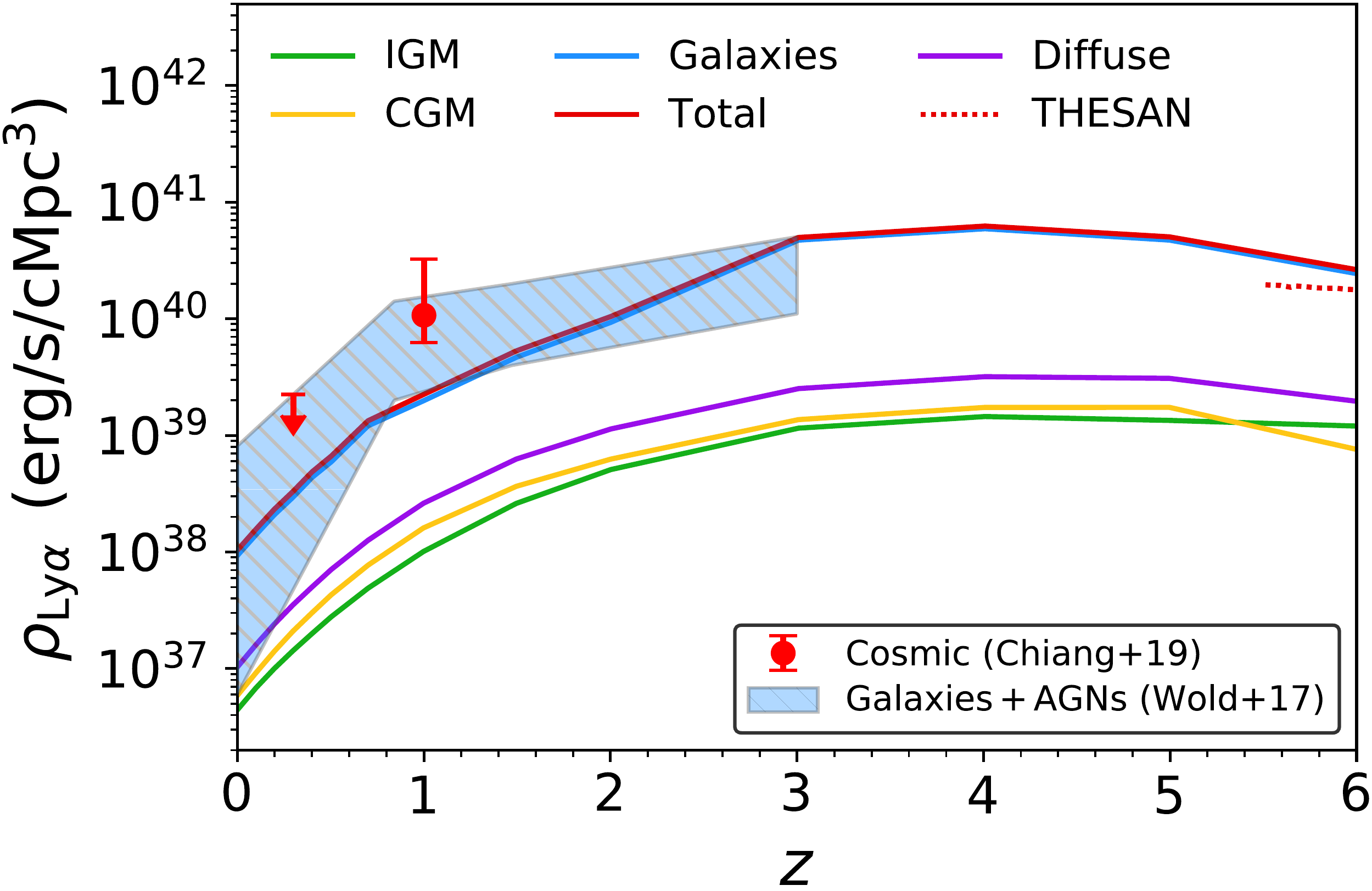}

\caption{The evolution of comoving $\rho_{\rm Ly\alpha}$. $\rho_{\rm Ly\alpha}$ increases with $z$ and peaks around $z \approx 4$, then decrease with $z$. The panel shows the $\rho_{\rm Ly\alpha}$ from the TNG100-1 and THESAN-1 data and compares it with recent observational results. The "Total" in this plot is defined as the sum of IGM, CGM, and Galaxies.}
\label{fig6}
\end{figure*}

In Figure~\ref{fig6}, we compare our results with observations and previous work. The results indicate that Ly$\alpha$ emission primarily originates from galaxies at any redshift. Even at the peak epoch ($z\sim4$) where diffuse emission is strongest, the diffuse gas accounts for at most $\sim16\%$ of the total Ly$\alpha$ luminosity density, much less than the contribution from galaxies. Only under conditions when galaxy Ly$\alpha$ output is diminished or diffuse output enhanced might the diffuse gas rival the galaxies. Our findings are consistent with those of \citep{2017Wold}, who found that the total Ly$\alpha$ luminosity density at $z \approx 0$ is dominated by galaxies. Notably, the galaxy Ly$\alpha$ luminosity in our simulation at $z = 0-0.3$ agrees with their observation, confirming the empirical treatment we adopt for the escape of Ly$\alpha$ photons from galaxies. More broadly, our $\rho_{\rm Ly\alpha}$ results remain in good agreement with \citep{2017Wold} across $z < 3$. However, compared to \cite{2019Chiang}, which reports the total $\rho_{\rm Ly\alpha}$ at $z = 0.3$ and $z = 1$ using {\it GALEX} data through intensity mapping, our predictions are approximately $7$ times and $5$ times smaller. There are a few possible explanations for this discrepancy. One is that the simulation may underpredict star formation or the number of Ly$\alpha$-emitting galaxies at $z\sim1$, or perhaps the escape fraction of Ly$\alpha$ from these galaxies is higher in reality than what we assumed. Another possibility is that the observational intensity mapping measurement suffers from systematic contamination (e.g., from continuum or foregrounds), which could have boosted the inferred Ly$\alpha$ signal.; indeed intensity mapping is challenging and the error bars are large. Our result might actually lie within the broad allowable range once uncertainties are considered. We also note that \cite{2019Chiang} included all Ly$\alpha$ emission in their measurement, so it is not straightforward to attribute the difference to one component or another. It could be that our simulation misses contributions from very faint galaxies below the resolution limit. For now, we highlight that our predicted Ly$\alpha$ intensity is somewhat below the currently observed level at $z=1$–3, but of the same order of magnitude.

We also present the $\rho_{\rm Ly\alpha}$ from the \texttt{THESAN} simulation in Figure~\ref{fig6}, covering the redshift range $z=6$ to $z=5.5$. Compared to the \texttt{TNG} results, the $\rho_{\rm Ly\alpha}$ in \texttt{THESAN} is systematically lower by a factor of 2-3. We attribute this difference primarily to the distinct radiation treatment: \texttt{THESAN} self-consistently solves the radiative transfer equations, whereas \texttt{THG} adopts a spatially uniform, tabulated UV background. This more accurate modeling of the local radiation field in \texttt{THESAN} probably leads to lower ionization rates in certain regions where the ionization was overionized in the \texttt{TNG} and consequently reduced the Ly$\alpha$ emission.

We estimate the observed upper limits of the IGM $\rho_{\rm Ly\alpha}$ by subtracting the integrated galaxy $\rho_{\rm Ly\alpha}$ from the total $\rho_{\rm Ly\alpha}$ values provided by \cite{2017Wold} and \cite{2019Chiang}. Although the total $\rho_{\rm Ly\alpha}$ luminosity density lies within the predicted range of integrated galaxy contributions, we adopt a conservative approach by subtracting the lower bounds of the galaxy terms, thereby establishing upper limits for the IGM component. Using this method, we find that the IGM upper limit at $z=0.3$ is $2.17 \times 10^{39}\, \rm erg/s/cMpc^{3}$, which is approximately 65 times higher than our calculation result. At $z=1$, the estimated upper limit is $2.97 \times 10^{40}\, \rm erg/s/cMpc^{3}$, which is 120 times larger than our result. In other words, current observations are nowhere near sensitive enough to detect the diffuse Ly$\alpha$ background directly; they only provide upper bounds that are orders of magnitude above the predicted signal. This strongly implies that direct detection of the cosmic diffuse Ly$\alpha$ emission is extremely challenging with present facilities, consistent with the fact that no unambiguous detection has been reported yet for truly diffuse Ly$\alpha$ (outside of the environments of bright quasars or galaxies).

Due to the challenges posed by observational limits and the low surface brightness of the diffuse Ly$\alpha$ signal, direct detection is challenging. However, our results suggest that Ly$\alpha$ emission density would enhance as $z$ increases. Thus, searching for Ly$\alpha$ emission from diffuse gas may improve the probability of detection in a high $z$ universe.  

\section{Discussion}
\subsection{Exclusion of star-forming gas from the Ly\texorpdfstring{$\boldsymbol{\alpha}$}{α} budget}

In our analysis of Ly$\alpha$ emission from the "Galaxies" region, we include contributions from stars as well as $SFR=0$ gas via both collisional excitation and recombination. However, we do not include the Ly$\alpha$ emission from the gas with $SFR>0$, despite its potential to contribute significantly to the intrinsic Ly$\alpha$ luminosity. This is because, in cosmological simulations, the gas with $SFR>0$ is typically part of the dense interstellar medium, which is characterized by high hydrogen densities, abundant in dust and negligible velocity gradients. Ly$\alpha$ photons produced within these regions undergo resonant scattering with neutral hydrogen, leading to extremely long path lengths and a high probability of absorption by dust. Consequently, the Ly$\alpha$ escape fraction from star-forming gas is expected to be very low, often by orders of magnitude lower than that from stars or surrounding diffuse gas. In contrast, Ly$\alpha$ photons originating from stellar sources can escape more easily, especially when feedback processes such as stellar winds and supernovae create low-density channels that facilitate escape \citep{Gronke2016}.

\subsection{Resolution dependence of diffuse Ly\texorpdfstring{$\boldsymbol{\alpha}$}{α} emission}

\begin{figure}
\centering
\includegraphics[width=0.5\textwidth]{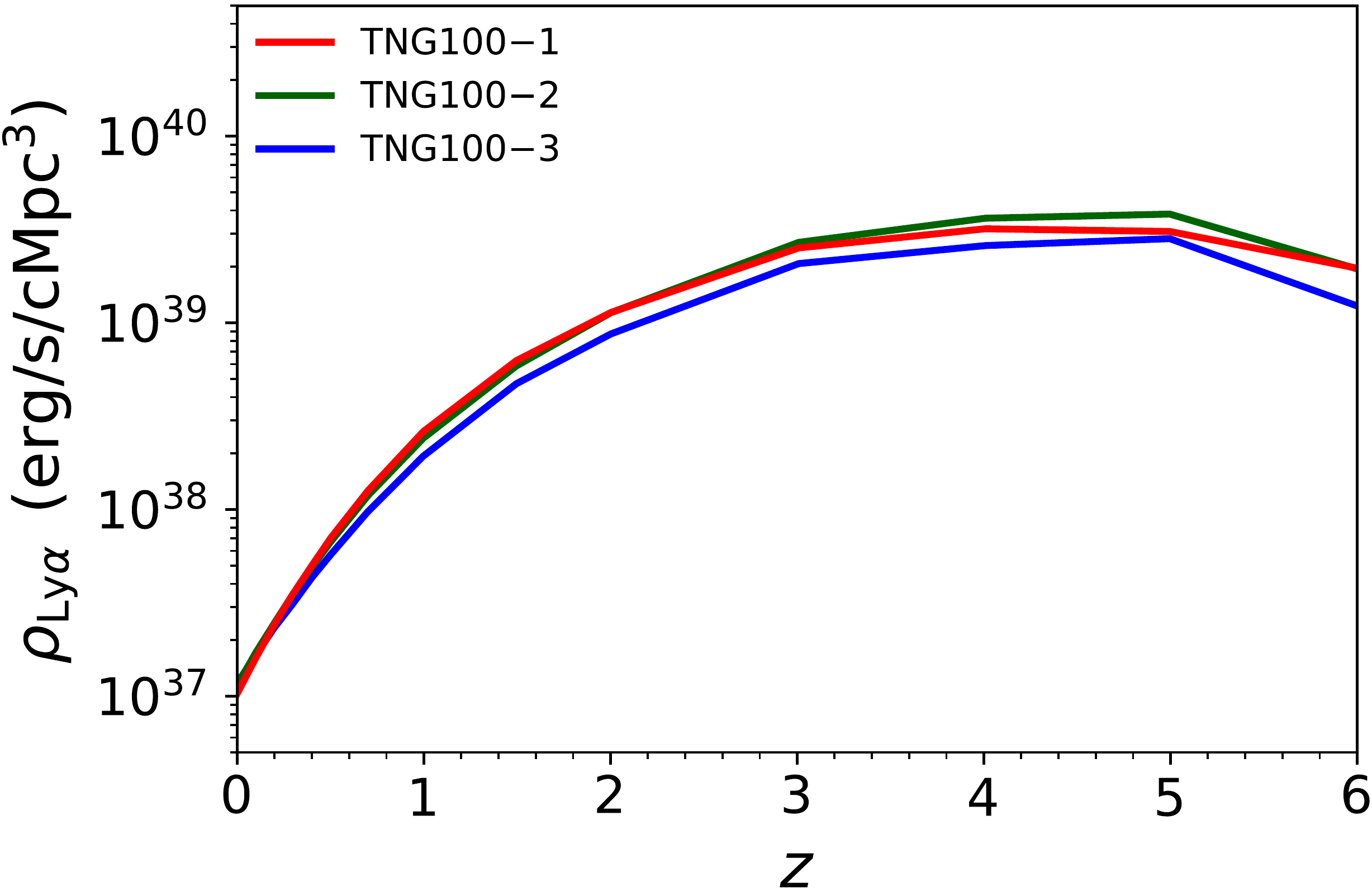}
\caption{The evolution of comoving diffuse $\rho_{\rm Ly\alpha}$ from the TNG100 with different resolutions; TNG100-1 (high), TNG100-2 (intermediate), TNG100-3 (low). These three profiles are consistent with each other. }
\label{fig7}
\end{figure}

We compare the evolution of the comoving diffuse $\rho_{\rm Ly\alpha}$ among TNG100-1, TNG100-2, and TNG100-3 in Figure~\ref{fig7}. Overall, the $\rho_{\rm Ly\alpha}$ values from different resolution runs are largely consistent with our definition of diffuse gas. The main difference arises during the transition from $z=6$ to $z=5$. As shown in Figure~\ref{fig5}, the $\rho_{\rm Ly\alpha}$ from recombination exhibits a sudden increase at $z=5$, corresponding to the point where the UV background is activated in the \texttt{TNG} simulations (starting at $z=6$). In TNG100-1, the contribution of recombination at $z=5$ accounts for approximately $66\%$ of the collision excitation in the IGM components. However, in lower-resolution runs (TNG100-2 and TNG100-3), the recombination luminosity increases sharply between $z=6$ and $z=5$, even exceeding the collisional excitation contribution. The lower resolution runs also show a steeper slope in the $\rho_{\rm Ly\alpha}$ evolution trajectory at $z=5-6$.  


\subsection{Comparison to previous theoretical studies}

Investigating the physical properties of diffuse gas is crucial for unraveling the mysteries of the physical conditions within the IGM. Therefore, studying the Ly$\alpha$ emission from these diffuse gases has a significant value. \citet{2020Elias} explore the potential detectability of large-scale filaments. To isolate filament particles, they exclude any particles identified by FoF within a halo and then extract a $5 \rm cMpc^{2}$ comoving region centered on the halo. When investigating the dependence of Ly$\alpha$ emission on physical conditions at $z=3$, they present temperature versus density phase-space diagrams. Compared with our results of phase diagrams, the morphology differs due to the different redshifts and variations in data coverage. In their study, the selected particles represent the gas surrounding a halo, depicting the physical properties of a smaller cosmic volume. Instead, we consider the diffuse gas within the simulation box with 100 cMpc on a side, including IGM in the FoF groups. Thus, our phase diagrams represent the general properties of diffuse gas. Their work also showed collisional excitation as the primary mechanism of diffuse Ly$\alpha$ emission, which is consistent with our findings. Therefore, collisional excitation could be the key mechanism driving diffuse Ly$\alpha$ emission at $z \geq3$.

We compare our results with those presented in Figure 3 of \citet{2021Witstok}, who investigated the prospects of observing Ly$\alpha$ emission from the low-density cosmic web. Although their results indicate that recombination dominates Ly$\alpha$ emission across most of the cosmic web, we find that in our simulations, collisional excitation is the dominant mechanism in both the IGM and the CGM. A key reason for this discrepancy lies in how the diffuse gas or IGM is defined in each study. In our work, diffuse gas includes all non-star-forming gas outside of galaxies, without imposing strict density or temperature thresholds. This includes both the IGM and the outer regions of halos typically classified as CGM. In contrast, \citet{2021Witstok} adopt a more conservative definition of the IGM, explicitly excluding gas above the self-shielding threshold density for UV background radiation. Following \citet{2013Rahmati}, they apply a cutoff around $n_{\rm H} \sim 10^{-2}$–$10^{-3}\ \rm{cm}^{-3}$, removing from their IGM classification any gas that would likely become self-shielded and optically thick. As a result, dense gas in halo outskirts, where collisional excitation is most efficient, is excluded from their IGM Ly$\alpha$ emission estimates. This naturally suppresses the collisional component and leads to a dominance of recombination emission in their analysis. The impact of this definition is especially apparent at lower redshifts, where much of the collisional Ly$\alpha$ emission in our model arises from dense CGM-like gas. If we were to impose a similar upper overdensity threshold (e.g., excluding gas with the density contrast, $\Delta > 100$), our diffuse Ly$\alpha$ luminosity density would drop substantially, particularly at $z \lesssim 3$. In that sense, our results can be seen as an upper bound on the diffuse Ly$\alpha$ signal, assuming even relatively dense outer-halo gas contributes and that its emission can escape. It shows the importance of density criteria for diffuse gas in predicting Ly$\alpha$. While \citet{2021Witstok}'s approach aims to isolate only the truly unshielded intergalactic medium, our broader inclusion captures a more complete picture of where Ly$\alpha$ photons originate, even if not all may be observable. We apply an escape fraction to approximate radiative losses, but we acknowledge this is a simplified treatment. More sophisticated models could refine this prediction in future work, such as including complete radiative transfer or explicitly excluding self-shielded clumps. Nevertheless, to estimate the maximum Ly$\alpha$ output from diffuse environments, our inclusive approach remains informative. Similar caution is also noted in \citet{Byrohl2023}, who emphasize that emission from the CGM can dominate and must be modeled carefully when forecasting the detectability of diffuse Ly$\alpha$ emission.

\subsection{Implications for Observations and Ly\texorpdfstring{$\boldsymbol{\alpha}$}{α} surveys}
Recent surveys using integral field unit (IFU) spectrographs on 8–10m class ground-based telescopes, such as MUSE and KCWI \citep{Bacon2010, Chen2021}, have revolutionized the study of faint Ly$\alpha$ emission, particularly in the CGM and IGM. These instruments have enabled the detection of extended Ly$\alpha$ nebulae around AGNs \citep{2017ASSL..430..195C} and filamentary Ly$\alpha$ emission in overdense galaxy environments, revealing the diffuse gas believed to fuel galaxy growth \citep{2019Umehata, Martin2019}. Surveys have also detected Ly$\alpha$ halos around star-forming galaxies extending tens of kpc, as well as filamentary structures around protoclusters and quasars \citep{2017Bacon, Chen2021, Martin2023}. These observations provide a direct view of the CGM and cosmic web, often revealing regions where collisional excitation dominates Ly$\alpha$ production, such as in high-density environments near massive halos.

For even fainter IGM Ly$\alpha$ emission in cosmic filaments, statistical stacking techniques could further enhance detectability, as demonstrated in studies of the Sunyaev–Zeldovich effect and X-ray emission \citep{2019A&A...624A..48D, 2019MNRAS.483..223T, 2020A&A...643L...2T}. Wide-field IFU surveys, such as HETDEX \citep{2021ApJ...923..217G, Davis2023}, or narrowband imaging projects like ODIN \citep{2024ApJ...962...36L}, could use stacking to uncover faint diffuse Ly$\alpha$ emission. This approach would leverage statistical power to reveal the diffuse component beyond individual detections.

Upcoming 30m class telescopes, such as the E-ELT, TMT, and GMT, will significantly enhance the detection of faint Ly$\alpha$ emission. With their large collecting areas and high-resolution IFU capabilities, these observatories are expected to probe the diffuse Ly$\alpha$ emission associated with the cosmic web, including fainter filaments beyond rare protocluster environments \citep{Byrohl2023}. Although our analysis focuses on the global luminosity density rather than spatially resolved brightness, the trends we observe suggest that with sufficient integration times and optimized observational strategies, these next-generation facilities may be able to access the lower end of the Ly$\alpha$ brightness distribution, especially through statistical or stacked analyses.

At even higher redshifts ($z \gtrsim 6$), Ly$\alpha$ emission becomes increasingly difficult to detect directly due to strong absorption by the neutral IGM. Nonetheless, space-based intensity mapping offers a promising alternative. The \textit{SPHEREx} mission, successfully launched in March 2025, is currently conducting an all-sky infrared spectroscopic survey and is expected to be sensitive to redshifted Ly$\alpha$ emission out to $z \sim 7$ (corresponding to $\lambda \sim 0.97, \mu$m). While foreground contamination from other emission lines (e.g., H$\alpha$, [OIII]) and zodiacal light remains a significant challenge, cross-correlation with 21cm maps or galaxy surveys may help isolate the Ly$\alpha$ signal. Our results suggest that although the diffuse Ly$\alpha$ luminosity density declines toward $z = 7$, it remains non-negligible if reionization heating has already occurred. Therefore, data from \textit{SPHEREx} could provide valuable constraints on the end stages of reionization and the thermal and ionization state of the early IGM.

\subsection{Caveats and Limitations of This Study}
While our analysis provides insight into the cosmic-averaged Ly$\alpha$ background, several limitations must be acknowledged, stemming from both the underlying simulation frameworks and our post-processing assumptions.

\textbf{Uncertainties in Subgrid Physics:} The \texttt{IllustrisTNG} and \texttt{THESAN} simulations rely on subgrid models to treat unresolved physics such as star formation, stellar and AGN feedback, and ISM thermodynamics. While these models are calibrated to reproduce large-scale galaxy statistics, they are not unique and introduce uncertainties. For example, the efficiency of AGN feedback strongly influences the thermodynamic state of the CGM. Overly strong feedback could eject or overheat gas, suppressing Ly$\alpha$ emission (since gas at $T \gg 10^6$K emits little Ly$\alpha$), while weak feedback could retain more cool gas, potentially boosting emission. Similarly, the ISM in \texttt{TNG} is modeled using an effective equation of state for gas above $n_{\rm H} \sim 0.1\rm cm^{-3}$, preventing the resolution of cold neutral clouds that might either emit or absorb Ly$\alpha$ photons. These choices influence our predictions for both the amount and spatial distribution of diffuse Ly$\alpha$.

\textbf{Ionization Modeling and the UV Background:}
\texttt{TNG} treats ionization in equilibrium under a spatially uniform UV background, neglecting local fluctuations and non-equilibrium effects. In reality, UV background intensity varies spatially and temporally, and dense regions may be more self-shielded than the approximate prescriptions used here. This could lead to underestimating the neutral hydrogen fraction and recombination rates in dense regions. In contrast, \texttt{THESAN} includes radiative transfer to capture these effects more accurately, but only up to $z = 5.5$. Beyond that, our results depend on equilibrium ionization models, which may not fully capture ionization states in partially self-shielded gas.

\textbf{Simplified Treatment of Ly$\alpha$ Escape:}
We do not perform Ly$\alpha$ radiative transfer and instead apply empirical escape fractions ($f_{\rm esc}$) to account for dust absorption and resonant scattering. For galaxies, this factor is redshift-dependent and calibrated to observations. For the CGM, we adopt the same $f_{\rm esc}$, but this is a simplification. The CGM may be more transparent (being lower density and more extended) or more opaque (if filled with small neutral clumps that trap Ly$\alpha$). Without complete radiative transfer, we treat CGM emission as an upper limit. Observationally, some Ly$\alpha$ photons from the CGM may scatter into the IGM or be absorbed by dust in outer disks. The true escape fraction of CGM Ly$\alpha$ remains uncertain, making this a key caveat in our results.

\textbf{Excluded Physical Sources:}
Our model includes only Ly$\alpha$ emission from collisional excitation and UV background driven recombination. We do not account for potentially minor sources such as X-ray heating (which can generate secondary ionizations), exotic decaying particles, or neutrino interactions. In addition, we do not explicitly include AGN line emission—although AGN contributions are partly captured through their influence on the UV background and in the \texttt{THESAN} radiative transfer field. If AGN were more numerous or luminous than assumed, they could contribute to the Ly$\alpha$ background via broad-line region emission or fluorescent illumination of nearby gas. However, their space density is low at $z \lesssim 6$, and stars are generally believed to dominate the ionizing photon budget during this epoch. Therefore, our assumption that stellar sources dominate the Ly$\alpha$ background at $z=0-6$ is consistent with current understanding.

\section{Conclusion} \label{sec:Conclusion}
We present a systematic investigation of LY$\alpha$ emission from diffuse cosmic gas, including CGM and IGM, based on post-processed data from the \texttt{IllustrisTNG} cosmological simulation suite, supplemented by the \texttt{THESAN} simulation run at high redshift ($z > 5.5$). We identified diffuse gas by utilizing the FoF and Subfind algorithms, and computed its Ly$\alpha$ luminosity from collisional excitation and recombination processes based on local gas properties. Our main conclusions are summarized as follows:
\begin{itemize}
\item \textbf{Emission mechanism:} Collisional excitation dominates the Ly$\alpha$ production in diffuse gas across all redshifts, particularly in gas with $T \sim 10^4$--$10^5$ K and $n_{\rm H} \sim 10^{-4}$--$10^{-2}\ \rm cm^{-3}$, such as shock-heated CGM or filamentary IGM.
Recombination contributes primarily from ionized regions ($T\sim10^4\,K$). At $z \sim 6$, collisional emission exceeds recombination by over two orders of magnitude; by $z\sim3$, the contribution from recombination increases to $\sim 35\%$, though collisional processes remain dominant.
\item \textbf{IGM vs CGM:} We find that Ly$\alpha$ emission below $n_{\rm H} \sim 10^{-2}\ \rm cm^{-3}$ is dominated by the IGM, while the CGM becomes the main contributor at higher densities. The CGM contributes 1.2–1.6 times more diffuse Ly$\alpha$ than the IGM throughout most of cosmic time. A prominent horizontal feature at $T \sim 10^4$ K in the density-temperature-emission phase diagrams arises from the balance between UV background photoheating and cooling inefficiencies, which thermally traps gas in this regime.
\item \textbf{Diffuse vs galaxy emission:} Galaxies dominate the total Ly$\alpha$ luminosity density at all redshifts. Even after accounting for empirical escape fractions, the stellar component consistently contributes over 90\% of Ly$\alpha$ emission, with diffuse gas (IGM+CGM) contributing $\lesssim$10–20\%. Therefore, observationally separating the faint diffuse background from the dominant galaxy signal remains a major challenge.
\item \textbf{Redshift evolution:} The diffuse Ly$\alpha$ luminosity density peaks around $z\sim4$, driven by active structure formation, efficient collisional processes, and a strong UV background. At $z\sim6$, we see a brief enhancement due to reionization heating, with the IGM temporary exceeding the CGM in Ly$\alpha$ output. After $z\sim4$, the diffuse component decreases sharply, by $z=0$ it is 2–3 orders of magnitude fainter than its peak. This evolution echoes that of the global star formation rate, and reflects the cooling and dispersion of the warm baryonic medium.
\item \textbf{Observational comparison and implications:} Our predictions for total Ly$\alpha$ luminosity density (galaxies + diffuse gas) broadly agree with available observations at $z=0$ and $z=3$, but fall below some intensity mapping estimates at $z=1-2$. This discrepancy may be due to systematic uncertainties in observations, or missing contributions from unresolved galaxies or AGN. Importantly, our predicted diffuse Ly$\alpha$ signal lies well below the current observational upper limits, particularly at $z=1-3$, suggesting that there is no strong contradiction. Future intensity mapping or cross-correlation techniques may be able to detect this elusive component.
\item \textbf{Prospects for detection:} While the diffuse Ly$\alpha$ background is intrinsically faint, it becomes relatively more prominent at higher redshift. Statistical techniques such as stacking or cross-correlation with galaxy density fields, especially using upcoming 30-m class telescopes (e.g. E-ELT, TMT, GMT), may reveal this signal. Our results offer a quantitative baseline for the expected strength and origin of this diffuse emission, which traces the warm, shock-heated cosmic web outside galaxies.
\end{itemize}
In summary, our study provides a detailed theoretical prediction for the cosmic diffuse Ly$\alpha$ background from IGM and CGM, clarifying its redshift evolution, physical origin, and detectability. While challenging to observe directly, this faint emission contains valuable information about the distribution, thermal state, and ionization history of baryons in the universe. As next-generation facilities come online, they will begin to probe this "cosmic Ly$\alpha$ glow," opening a new window into the low-density universe that fuels galaxy formation.


\section*{Acknowledgements}

This research is supported by the National Science and Technology Council, Taiwan, under grant No. MOST 110-2112-M-001-068-MY3, NSTC 113-2112-M-001-028-, NSTC 111-2112-M-001-090-MY3, and the Academia Sinica, Taiwan's career development awards under grant No. AS-CDA-111-M04 and AS-CDA-113-M01. This research was supported in part by grant NSF PHY-2309135 to the Kavli Institute for Theoretical Physics (KITP) and grant NSF PHY-2210452 to the Aspen Center for Physics. KC acknowledges the support of the Alexander von Humboldt Foundation and Heidelberg Institute for Theoretical Studies.
Our computing resources were supported by the National Energy Research Scientific Computing Center (NERSC), a U.S. Department of Energy Office of Science User Facility operated under Contract No. DE-AC02-05CH11231 and the TIARA Cluster at the Academia Sinica Institute of Astronomy and Astrophysics (ASIAA).




\bibliographystyle{mnras}
\bibliography{refs} 








\bsp	
\label{lastpage}
\end{document}